\documentclass[aps,prl,superscriptaddress,reprint,showpacs,preprintnumbers,amsmath,amssymb]{revtex4-1}

\usepackage{latexsym}
\usepackage{amsmath}
\usepackage{amssymb}
\usepackage{epsfig}
\usepackage{dcolumn}% Align table columns on decimal point
\usepackage{bm}% bold math
\usepackage{graphicx}
\graphicspath{{figures/}}
\usepackage{listings}
\usepackage{comment}
\usepackage{color}
\usepackage{wasysym}
\usepackage{mathtools}
\usepackage{color}
\begin{document}

%\preprint{APS/123-QED}
%\pacs{47.55.D-, 47.32.ck, 47.20.Ma, 47.20.Dr}
% http://www.aip.org/pacs/pacs2010/individuals/pacs2010_regular_edition/index.html
% 47.55.D-: Drops and bubbles
% 47.32.ck:	Vortex streets
% 47.20.Ma: Interfacial instabilities (e.g., Rayleigh-Taylor)
% 47.20.Dr: Surface-tension-driven instability

\title{Conical-focusing:  Mechanism for singular jetting from collapsing drop-impact craters}

\author{Yuan Si \surname{Tian}}
\thanks{These two authors contributed equally}
\affiliation{School of Construction Machinery, Chang'an University, Xi'an, 710064, China}
\affiliation{Division of Physical Sciences and Engineering,
King Abdullah University of Science and Technology (KAUST),
Thuwal, 23955-6900, Saudi Arabia}

\author{Zi Qiang \surname{Yang}}
\thanks{These two authors contributed equally}
\affiliation{Division of Physical Sciences and Engineering,
King Abdullah University of Science and Technology (KAUST),
Thuwal, 23955-6900, Saudi Arabia}

\author{Sigurdur T. \surname{Thoroddsen}}
\email{\textcolor{black}{sigurdur.thoroddsen@kaust.edu.sa}}
\affiliation{Division of Physical Sciences and Engineering,
King Abdullah University of Science and Technology (KAUST),
Thuwal, 23955-6900, Saudi Arabia}
\date{\today}

\begin{abstract}
Fast microjets can emerge out of 
liquid pools from the rebounding of drop-impact craters, or when a bubble bursts at it surface.  The fastest jets are the narrowest and are a source of aerosols both from the ocean and a glass of champagne, of importance to climate and the olfactory senses.  The most singular jets\textcolor{black}{, which have a maximum velocity of 137$\pm$4 m/s and diameter of 12 $\mu$m under reduced ambient pressure,} are produced when a small dimple forms at the crater bottom and rebounds without pinching off a small bubble.  The rebounding of this dimple is purely inertial but highly sensitive on initial conditions.  High-resolution numerical simulations reveal a new focusing mechanism, which drives the fastest jet within a converging conical channel, where an entrained air-sheet provides effective slip at the outer boundary of the conically converging flow into the jet.  This configuration bypasses any viscous cut-off of the jetting speed and explains the extreme sensitivity observed in detailed experiments of the phenomenon.  
%The simulations also do not require arbitrary free-surface reconnections which are required when the dimple pinches off.
\end{abstract}

\maketitle

\textcolor{black}{%{\it Introduction:}  
The fundamental singularities of free-surface flows are usually associated with pinch-off or coalescence \cite{Eggers1997,Eggers2015singularities}. 
The pinch-off of an inviscid liquid thread shows self-similar capillary-inertial dynamics \cite{DayHinchLister1998,Keim2006,Schmidt2009memory}, while a bubble pinches off in a purely inertial motion and the surface tension becomes irrelevant during the final stage of collapse \cite{Taborek2005scaling,Thoroddsen2007,Eggers2007}.  
The fine jetting observed from the bottom of a rebounding free-surface crater, generated from critical Faraday waves or bursting bubbles, has also been suggested to arise from a capillary-inertial singularity \cite{Zeff2000,Duchemin_2002,Das2008,Lai2018,Deike2018}. 
However, experiments show this not to be correct during the final stages of drop-impact crater collapse, where the dimple dynamics are purely inertial \cite{Thoroddsen2018,Yang2020}.  
The accelerating collision of the cylindrical walls of the dimple \cite{Oguz1990bubble,Pumphrey1990entrainment,Prosperetti1993impact} might be expected to create large impulsive pressure, akin to the spherical collapse of a cavitation bubble.
However, somewhat counter-intuitively the fastest jets occur not when the dimple pinches off a small bubble, but rather when the narrowest dimple retracts vertically just before pinch-off.
Thoroddsen {\it et al.} \cite{Thoroddsen2018} used ultra-high-speed imaging to show that this occurs without any curvature singularity at the tip of the dimple.}

\textcolor{black}{Questions remain: how does the cross-over from the radial collapse to vertical jetting occur?  What is the largest jet velocity and is there an upper bound? Furthermore, why is this speed so sensitive to the boundary or initial conditions? 
Herein we will reveal a conical jetting mechanism which can answer all of these questions.}

\begin{figure}[]
  \centering
  \includegraphics[width=1.0\linewidth]{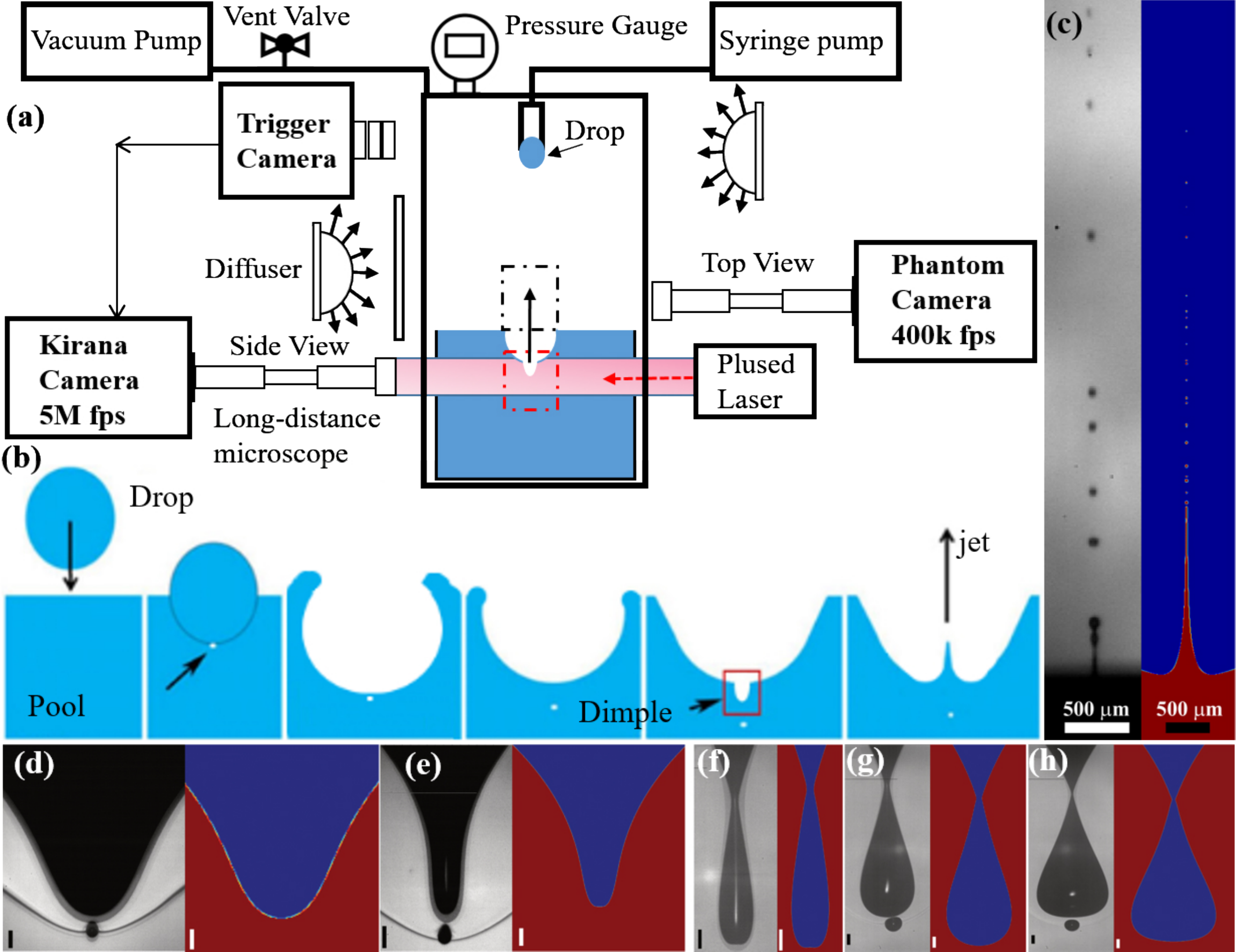}\vspace{-0.03in}\\%ziqiang  {figure_v1.png}
	\caption{(a) Experimental setup used to study crater collapse inside a vacuum chamber. 
	(b) Sketch of typical drop impact crater evolution, bubble entrapment, rebound and jetting.
	\textcolor{black}{(c) Overall singular jet shape and small fine droplets shooting from the drop-impact crater.}
	(d-h) Comparison between experiments and numerics of different dimple geometries and pinch-offs, under identical conditions with liquid viscosity $\mu=7.3$ cP and $D=3.64$ mm, for (d) $U =$ 1.30 m/s, $We =$ 100, $Fr =$ 49; (e) $U =$ 1.38 m/s, $We =$ 113, $Fr =$ 55; (f) $U =$ 1.45 m/s, $We =$ 127, $Fr =$ 59; (g) $U =$ 1.49 m/s, $We = $ 134, $Fr = $ 63; (h) $U =$ 1.54 m/s, $We = $ 143, $Fr = $ 67. The \textcolor{black}{unmarked} scale bars are 50 $\mu$m.
	}
  \label{Fig_1}
  \vspace{-0.2in}
\end{figure}

% \cite{Yarin2006,JosserandThoroddsen2016}.
\textcolor{black}{The dynamical importance of the surrounding air in free-surface flows was conclusively demonstrated by Xu {\it et al.} \cite{XuZhangNagel2005},
who showed that impact splashing can be suppressed by reducing the ambient pressure. The observations of micro-bubbles near the most singular jetting, from drop-impact craters, suggested air-entrainment or even possible cavitation \cite{Thoroddsen2018,Tran2016}.  
This motivated us to conduct a set of experiments, under reduced ambient pressures, to pinpoint the role of the gas on the jetting.  This also serves a second purpose to correct the jetting speed for the air-drag experienced by the tip of the jet before it emerges out of the crater.  Below we show how air-sheets play an unexpected role in promoting the fastest jets. 
}
%Mandre {\it et al.} \cite{Mandre2009} proposed to explain this by suggesting the air-layer makes the drop skate along the surface without making contact. 

\begin{figure}
  \centering
      \includegraphics[width=1.0\linewidth]{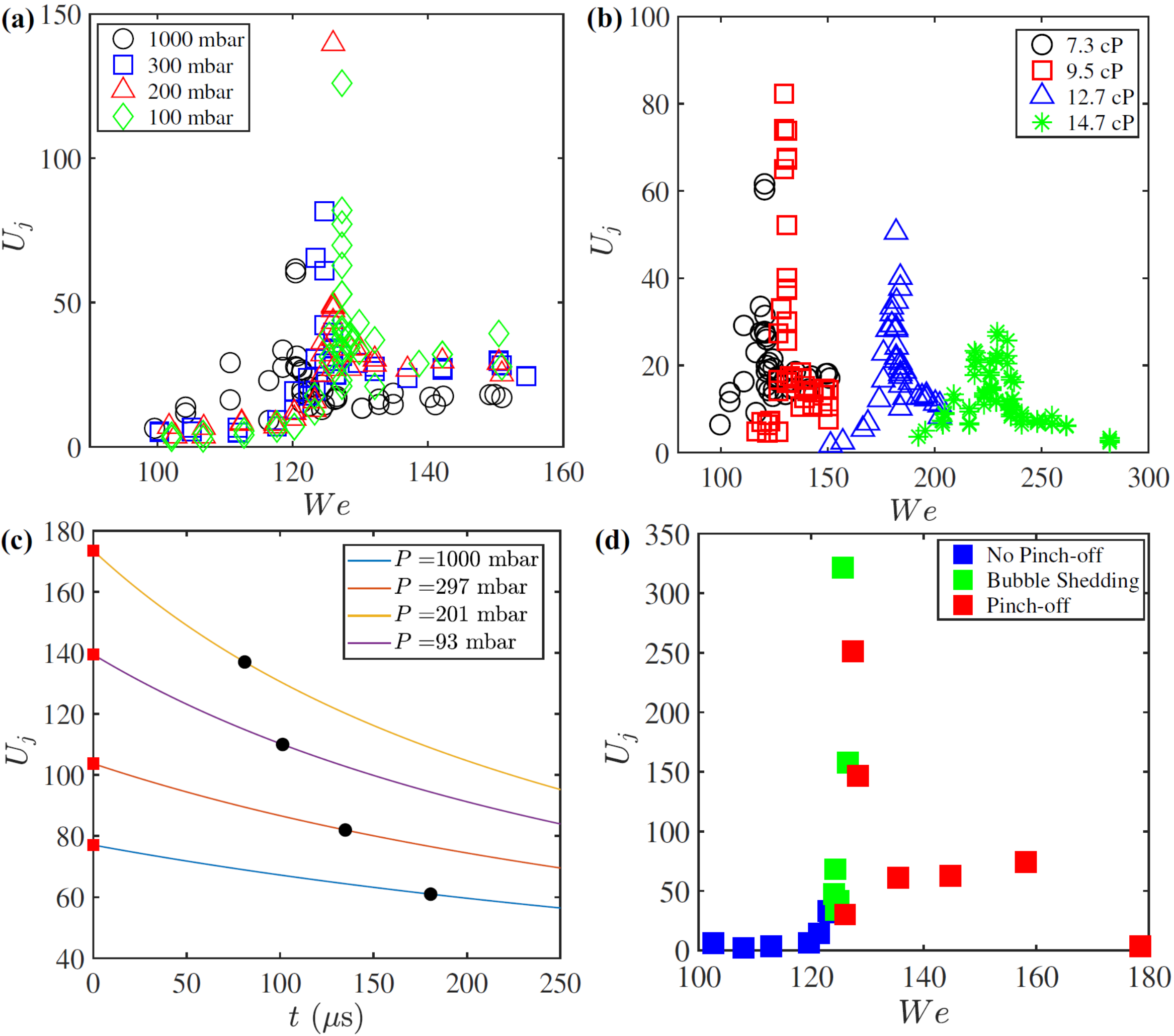}\vspace{-0.1in}\\
  \caption{\textcolor{black}{Jet characterization from the experiments (a-c) and numerical simulations (d).  (a) Jetting speed $U_j$ vs impact Weber number for different ambient pressures, for liquid viscosity $\mu=7.3$ cP.  (b) $U_j$ for different liquid viscosities, at 1 atmosphere pressure.
  (c)  Extrapolation of jetting velocity to emergence, from the observed speed (black circles) coming out of the crater, accounting for the air drag.
  (d) Results from Gerris simulations for $\mu=7.3$ cP at atmospheric pressure}} 
  \label{Fig_2}
\end{figure}

\textcolor{black}{The experimental configuration is shown in Fig. \ref{Fig_1} and is similar to that used in our previous work 
\cite{Thoroddsen2018,Yang2020,Li2017},
while now the impacts are performed inside an acrylic vacuum chamber.
The dimple evolution is viewed through the glass pool, 
using a long-distance microscope (Leica Z16 APO) with adjustable magnification and aperture, at pixel resolution down to 1.1 $\mu$m/px.
The rapid motions demands observations with an ultra-high-speed video camera (Kirana-05M, Specialized Imaging, Tring UK), which acquires 180 frames at up to 5 million fps
with full-frame resolution of $924\times768$ px irrespective of the frame rate used.  Illumination is provided by 180 diode-lasers (SI-LUX640), one for each frame.  The pulse-duration is between 30 - 170 ns to eliminate any motion smearing.  A second video camera captures the jet rising out of the crater to measure its speed and width.  Triggering is electronic, started when the falling drop cuts the light to a dedicated line-sensor.  The drop has a fixed diameter $D\simeq 3.64$ mm, using water/glycerin mixtures, to cover a small range of viscosities of $\mu = 7.3$, 9.5, 12.7 cP \& 14.7 cP, impacting the pool at velocity $U$. The surface tension $\sigma \simeq $ 68  mN/m.
The impacts are characterized by Weber and Froude numbers, $We =\rho D  U^2/\sigma, \;  Fr= U^2/(gD)$,
where we use the effective diameter $D = ({D_H}^2D_V)^{1/3}$ at impact, as monitored by the upper camera. \textcolor{black}{$D_H$ and $D_V$ stand for the horizontal and vertical diameter of the drop at impact.}}

\textcolor{black}{The axisymmetric numerical simulations use the open-source {\it volume-of-fluid} Gerris software \cite{Popinet2009,Popinet2018}, which solves the incompressible Navier-Stokes equations in both the gas and liquid, using exact densities, viscosities and surface tension as in the experiments.  To capture the fine air-sheets the code uses extreme grid refinement at the free surface, starting at refinement levels of 12, increasing to 16 or 18 close to the start of jetting (see Supplemental Material \cite{Supp}).}
\textcolor{black}{The large density ratio between the two phases limits the simulations to the ambient pressure case. We remove the initial entrapped bubble during the first contact of the drop with the pool, to reduce the computational burden to allow the extreme refinement during the dimple dynamics.  We note that this central bubble is absent during singular jetting from bursting bubbles and supercritical Faraday waves \textcolor{black}{\cite{Zeff2000, Das2008}}.}
% [Refs].
% Longuet-Higgins Reference Longuet-Higgins1983; 
% Zeff et al. Reference Zeff, Kleber, Fineberg and Lathrop2000; 
% Das & Hopfinger Reference Das and Hopfinger2008

\begin{figure}[t!]
  \centering
    \includegraphics[width=1.0\linewidth]{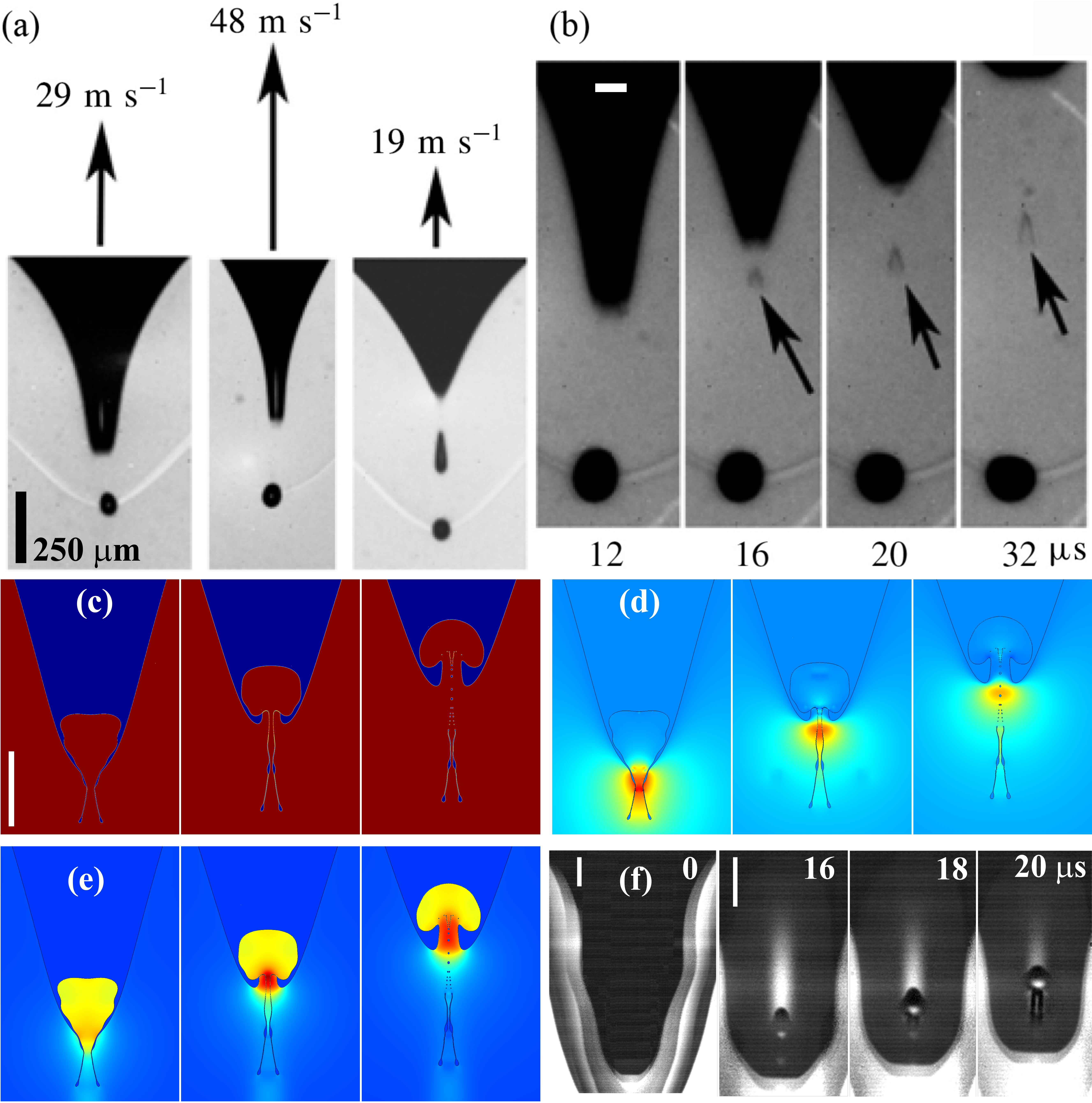}\vspace{-0.08in}\\
	\caption{(a) Dimple shapes and the corresponding jet speeds. %\textcolor{blue}{The scale bar is 250 $\mu$m.} 
	(b) The wedge-shaped microbubbles shed from the tip of the dimple, modified from Thoroddsen {\it et al.} \cite{Thoroddsen2018}. 
	The small bubble in all of the frames comes from the initial air-disc entrapped under the drop when it first hits the pool surface \cite{Peck1994,Thoroddsen2003,Jian2020}.
	\textcolor{black}{This bubble is removed in the numerical simulations.}
	(c-e) Simulation results from the second-highest green square in Fig. 2(d), showing the broad jet tip shape (c), pressure field (d) and vertical velocity (e).
	Images are spaced by 0.76 $\mu$s (f) Example experimental results with similar thick jet tip visible inside the dimple. % 	, $dt=0.2\; \mu$s??
	The \textcolor{black}{unmarked} scale bars are 50 $\mu$m.
	}
  \label{Fig_3}
\end{figure}

%%%%%%%%%%%%%%%%%%%%%%%%%%%%%%%%%%%%%%%%%%%%%%%%%%%%%%%%%%%%%%%%%%%%%%%%%%%%%%%%%%%%%%%%%%%%%
%   RESULTS
%%%%%%%%%%%%%%%%%%%%%%%%%%%%%%%%%%%%%%%%%%%%%%%%%%%%%%%%%%%%%%%%%%%%%%%%%%%%%%%%%%%%%%%%%%%%%

\begin{figure*}
  \centering
	  \includegraphics[width=0.95\textwidth]{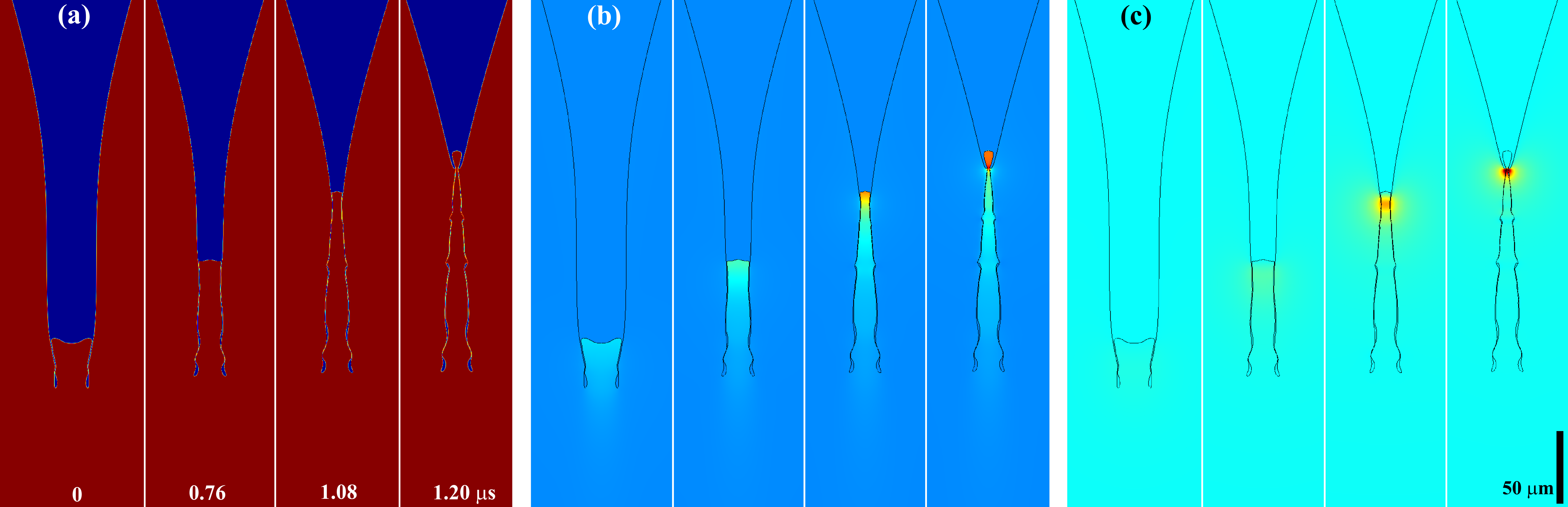}\vspace{-0.1in}\\   % 100 um = 95 px
  \caption{The cone-focusing mechanism driving out the fastest narrow jet, at $U_j=320$ m/s, emerging under the following conditions:  $D=3.64$ mm, $U=1.435$ m/s, giving $We= 125$, $Fr=58$, for $\mu=7.3$ cP and at atmospheric pressure.  (a) interface shape; (b) vertical velocity; (c) dynamic pressure.  This entire sequence lasts %for normalized time of $t^*=9.5\times 10^{-5}$, which corresponds to 
  for real time of $t=t^* L/U = 1.20\; \mu$s.}  
  %The liquid is shot out of a ``barrel'' which narrows in the direction of motion, as well as in time.  The phase of these two motions determines the final diameter and velocity.}
  %at ($P=4.3$ kPa), $V= 2.27$ m/s, $R_b= 3.66$  mm 
%  Frames are shown at times relative to the fastest vertical retraction of the dimple as in \cite{Thoroddsen2018,Yang2020}.}
  %The scale bar is 100 $\mu$m long.}
  \label{Fig_4}
\end{figure*}

\textcolor{black}{Figure \ref{Fig_1}(c-g) compare experiments and simulations of a series of close-up dimple shapes at the bottom of the drop-impact craters near the singular collapse, as well as for dimple pinch-offs.  The agreement is quite good, with the pinched off volume slightly smaller in the experiments, but keep in mind that in the experiments the gas compresses by the larger {\it collision} pressure, while the simulations are incompressible. 
The corresponding jet velocities in Fig. \ref{Fig_2}(a,b,d) show a familiar trend with isolated peaks occurring over a very narrow range of $We$ numbers \cite{Zeff2000,Thoroddsen2018,Yang2020,Michon2017}.  This sensitivity to impact conditions has branded the singular jets as being {\it ``barely reproducible''} \cite{Michon2017}.  Reducing the ambient pressure shifts this critical $We_c$ from 121 to 127 and we observe a general trend of higher jetting velocity for lower air pressure, reaching the maximum measured velocity of 137\textcolor{black}{$\pm$4} m/s.}   

\textcolor{black}{Increasing the liquid viscosity shifts $We_c$ to higher values ($We_c = 182$ for $\mu = 12.7$ cP), but the character of the curves remains the same, until $\mu = 14.7$ cP, where the peak jetting velocity is much weaker and the peak becomes broader.  Viscosty below this value does not seem to control the nature of the jetting process, but will rather modify the phase and amplitude of the capillary waves traveling down the crater towards the bottom dimple \cite{Yang2020,Liow2001splash}.}

The maximum jetting velocity for reduced ambient pressure, in Fig. \ref{Fig_2}(b), can partly be explained by the reduced air-drag acting on the tip of the jet, which often pinches off before it emerges out of the crater, as these droplets are traveling orders of magnitude faster than their terminal velocity \cite{Thoroddsen2012}.  In the inertial regime, this drag-force scales linearly with the air-density.  In Fig. \ref{Fig_2}(c) we have accounted for this drag and extrapolated back to the emergence position.  This suggests the largest initial jet velocity of $U_j \simeq 175$ m/s.  The experiments do not show measurably narrower dimples as ambient pressure is reduced.
%(Is the dimple narrower for the reduced pressure??)

\textcolor{black}{Figure \ref{Fig_2}(d) shows $U_j$ from the Gerris simulations, reproducing the very narrow range of $We_c\simeq 125$,
where the largest jetting velocities are observed.  The fastest speed of 320 m/s occurs without dimple pinch-off, but when micro-bubbles or air-sheet is pulled out of the bottom corner of the dimple, which plays a crucial role in the below proposed jetting mechanism.
But why is there such a large range of jet velocities for the green data points without pinch-off?} 
\textcolor{black}{Figures \ref{Fig_3}(c-e) shows the shape at emergence for the second-highest speed jet which has a broad mushroom-like tip.  
Such broader tips are visible, inside the dimple, in some of the experiments (Fig. \ref{Fig_3}f).  This configuration leaves a wedge-shaped air-sheet similar to that which has been observed in isolated experiments, Fig. \ref{Fig_3}(b).}

\begin{figure}[t!]
  \centering
    \includegraphics[width=0.68\linewidth]{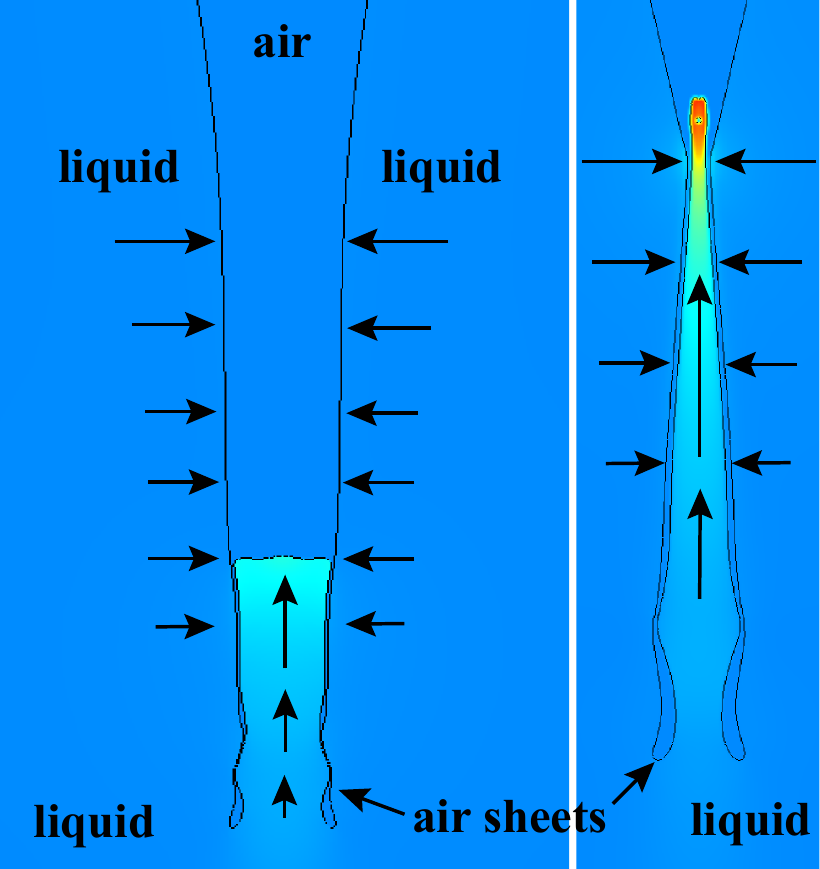}\vspace{-0.08in}\\   
	\caption{\textcolor{black}{Schematic of the new focusing mechanism. The arrows indicate the local velocity of the liquid. This highlights the decoupling between the radial collapse of the dimple wall and the vertical jetting within in the inner cone, with the two regions separated by a thin continuous air layer. The free surfaces are taken from the actual simulation in Fig. 4.  The color indicates the vertical velocity. The original width of the dimple is here
 $\sim 40\; \mu$m.}
	}
  \label{Fig_5}
\end{figure}

\textcolor{black}{\textcolor{blue}{ Figure} \ref{Fig_4} reveals the mechanism driving the fastest jets, during the rapid vertical ``retraction'' of the bottom of the cylindrical dimple.  An air-sheet is pulled from the corner of the air-dimple and the liquid is forced up into the cylindrical air cavity, while maintaining a thin air-sheet between the radially converging dimple and the vertically moving squeezed jet.  Even though the air is thin it provides a free-slip boundary protecting the jet from viscous stress.
The liquid is thereby shot out of a ``barrel'' which narrows in the direction of motion, as well as in time.  The phase of these two motions, radial and axial, determines the final diameter and velocity of the jet, as it emerges out of the dimple.
This mechanism is sketched in Fig. \ref{Fig_5}.  Yang {\it et al.} \cite{Yang2020} showed that the fastest jets are generated by the narrowest dimple which does not pinch off a bubble, i.e. what they call the {\it telescopic dimple}.
Now the reason for this is clear, as the walls must confine, accelerate and direct the fine jet.  This follows the simplest possible flow, i. e. that of accelerating towards an axisymmetric sink.}

\textcolor{black}{Simple volume conservation of a squeezed liquid cylinder inside the dimple, predicts a consistent jet speed
\[
U_j = \frac{1}{\pi R^2} \times 2\pi R h \frac{dR}{dt} = \frac{2h}{R}   \frac{dR}{dt} \sim 300\; m/s,
\]
when we use the height of the dimple $h=100\; \mu$m and radial collapse at 15 m/s, estimated from the dimple diameter reducing from 40 to 10 $\mu$m in $\sim 1 \; \mu$s.
The local Reynolds number of the $d_j=10\; \mu$m jet is $Re_j = d_j U_j/\nu \simeq 400$,
suggesting viscosity has not limited this velocity.
With prefect alignment, we don't see why these jets cannot be even thinner and faster}\textcolor{black}{, i.e. when the radial collapse has reached an even smaller diameter at the exact time of jet emergence.}

\textcolor{black}{The most unexpected aspect of this mechanism is the effective slip provided by the conical air-film at the outer edge of the base of the jet, which brings to mind other geometries where air-films play a key role, e.g. the drag reduction from the vapor-layer on a free-falling Leidenfrost sphere, as has been realized over a range of viscosities \cite{Vakarelski2016}; leaping shampoo also glides on a 0.5 $\mu$m air film \cite{Lee2013}. 
This neutralizing of the viscous shear may explain why our jets \cite{Thoroddsen2018,Yang2020} are much faster than those in the related dimple-geometry from bursting bubbles at a pool surface \cite{Walls2015,Bird2018,Ganan-Calvo2017, Ganan-Calvo2018,Gordillo2018,Gordillo2019,Seon2017}.   The suggested bubble-bursting theories include a viscous length-scale, through the Ohnesorge number, but the proposed scaling does not work for our jet velocities, see section 4 in Yang {\it et al.} \cite{Yang2020}.  The decoupling of the radial and axial motions will not be captured by flows demanding continuous velocity fields \cite{Gekle2009}.}

\textcolor{black}{How realistic are the Gerris simulations?  They are known to capture the finest details of splashing \cite{Thoraval2012} and here they closely reproduced  the dimple shapes and the critical Weber number for singular jets $We_c \simeq 125$, with any deviation likely arising from the drop shape, which is kept spherical in the simulations, while it oscillates slightly in the experiments. Keep in mind that {\it volume-of-fluid} simulations have a finite cut-off where a small bubble or air-film is eliminated and smoothed out.  The final dynamics of a neck pinch-off is therefore achieved artificially.  On the other hand for the singular cases herein, they occur without pinch-off bypassing this problem, but also requiring extreme grid refinement.   The air-films in Fig. \ref{Fig_4} are $\sim 1 \; \mu$m thick, which is three %\textcolor{black}{(four or three??? the dimple size should be around 5 mm.)} 
orders of magnitude smaller than the crater size, presenting significant challenges to numerical studies, even in the axisymmetric case.   Comparison with experiments shows qualitatively similar results, with the fastest jet velocities, after accounting for air drag, at up to 175 m/s.   The larger simulated velocity of 320 m/s, can obviously arise from the extreme sensitivity to boundary conditions, or can be affected by two properties of the numerical algorithm:  First, the gas and liquid are both incompressible, whereas the air-dimple is known to compress measurably from the high dynamic pressure and then emit sound-waves through the Minneart mechanism \cite{Pumphrey1990entrainment,Prosperetti1993impact,Thoroddsen2018, Phillips2018}.  Secondly, our simulations impose axisymmetry, which prevents the air-film from rupturing in the azimuthal direction, or from its edge \cite{Thoraval2020}. \textcolor{black}{The compressibility effects can be prominent as the Mach number $Ma=U_j/c$ of the maximum experimental jet speed is $Ma\simeq 0.40$, where $c$ is the speed of sound which is independent of ambient pressure.  This may contribute additional drag on the jet tip.  The numerics are incompressible, but suggest even larger $U_j$ are possibly, reaching conditions of relevance to forced jets for applications \cite{Thoroddsen2009,Tagawa2012,Rohilla2020}.}  The stability of the air-sheets and jet motion inside the dimple could in the future be studied with x-ray imaging with sub-microsecond time-resolution, which is beyond current capabilities.}

%\textcolor{black}{The fine jetting occurs not gradually, but at very narrow range of impact conditions, that the phenomenon is ``barely reproducible''.  This raises the question whether this is due to the emergence of a new mechanism, different from the gradual changes observed in the Worthington jet with changes in impact conditions.  Our conclusion is yes, the entrainment of the air-sheets changes the phenomenon. The timing of the outer and inner flow become decoupled, explaining the extreme sensitivity.}  

%\textcolor{blue}{The driving pressure will scale as the local radial velocity when the dimple stops collapsing, i.e. $P \sim \delta_{dimple}^2$ ?  Therefore, the thinner the dimple the larger the driving pressure.  In addition to this the conical funnel closes up with time, forcing more liquid upwards through a narrower throat, further accelerating the tip of the jet.}

Herein, we have identified the missing link between radial dimple collapse and the fastest vertical micro-jetting.
The effective slip from the conical air-film decouples the radial and vertical motions and the freedom in the phase between the two motions explains the extreme sensitivity to initial conditions.  It furthermore explains the failure of viscous scaling for the fastest jets as well as their extreme velocities. 
In addition it suggests that even faster jetting could be induced. \vspace{-0.1in}\\
%Clearly, this mechanism is not universal, nor a progression from the self-similar dimple shown in Zeff {\it et al.} \cite{Zeff2000} ?}\vspace{0.05in}\\

\begin{acknowledgements}
{This study was supported by King Abdullah University of Science and Technology
(KAUST) under URF/1/3727-01-01 and BAS/1/1352-01-01.
Yuan Si Tian is also supported by the Fundamental Research Funds for the Central Universities, CHD (grant no. 300102252109).}\\
%\backsection[Author ORCID]{ 
%Y. S. Tian, ORCID: 0000-0002-9705-2995;\\ 
%Z. Q. Yang, ORCID: 0000-0002-7760-2996;\\
%S. T. Thoroddsen, ORCID: 0000-0001-6997-4311.

%Li acknowledges the Thousand Young Talents Program of China, the National Natural Science Foundation of China (Grants 11642019, 11772327 and 11621202) and Fundamental  Research   Funds  for  the  Central  Universities (Grant  WK2090050041).

\end{acknowledgements}

\end{document}